\documentstyle[11pt,aaspp4]{article}

\input{psfig.sty}

\begin{document}

\title{Large Torque Variations in Two Soft Gamma Repeaters}

\author{
Peter~M.~Woods\altaffilmark{1,2},
Chryssa~Kouveliotou\altaffilmark{2,3},
Ersin~{G\"o\u{g}\"u\c{s}}\altaffilmark{2,4},
Mark~H.~Finger\altaffilmark{1,2},
Jean~Swank\altaffilmark{5},
Craig~B.~Markwardt\altaffilmark{5}, 
Kevin~Hurley\altaffilmark{6} and
Michiel~van~der~Klis\altaffilmark{7}
}

\altaffiltext{1}{Universities Space Research Association; 
peter.woods@msfc.nasa.gov}
\altaffiltext{2}{National Space Science and Technology Center, 320 Sparkman Dr. 
Huntsville, AL 35805}
\altaffiltext{3}{NASA Marshall Space Flight Center}
\altaffiltext{4}{Department of Physics, University of Alabama in Huntsville, 
Huntsville, AL 35899}
\altaffiltext{5}{NASA Goddard Space Flight Center, Greenbelt, MD 20771}
\altaffiltext{6}{University of California at Berkeley, Space Sciences
Laboratory, Berkeley, CA 94720-7450}
\altaffiltext{7}{Astronomical Institute ``Anton Pannekoek'' and CHEAF,
University of Amsterdam, 403 Kruislaan, 1098 SJ Amsterdam, NL}

\begin{abstract}

We have monitored the pulse frequencies of the two soft gamma repeaters
SGR~1806$-$20 and SGR~1900$+$14 through the beginning of year 2001 using
primarily {\it Rossi X-ray Timing Explorer} Proportional Counter Array
observations.  In both sources, we observe large changes in the spin-down
torque up to a factor of $\sim$4, which persist for several months.  Using long
baseline phase-connected timing solutions as well as the overall frequency
histories, we construct torque noise power spectra for each SGR.  The power
spectrum of each source is very red (power-law slope $\sim-$3.5).  The torque
noise power levels are consistent with some accreting systems on time scales of
$\sim$1 year, yet the full power spectrum is much steeper in frequency than any
known accreting source.  To the best of our knowledge, torque noise power
spectra with a comparably steep frequency dependence have only been seen in
young, glitching radio pulsars (e.g.\ Vela).  The observed changes in spin-down
rate do not correlate with burst activity, therefore, the physical mechanisms
behind each phenomenon are also likely unrelated.  Within the context of the
magnetar model, seismic activity cannot account for both the bursts and the
long-term torque changes unless the seismically active regions are decoupled
from one another.

\end{abstract}


\newpage

\section{Introduction}

Shortly after the first radio pulsars were discovered in 1967 (Hewish et al.\
1968), it was found that the pulse frequency of the Crab (one of the early
pulsars) was decreasing steadily with time (Richards \& Comella 1969).  This
behavior matched very nicely with that expected of an isolated, magnetized
neutron star spinning down via magnetic braking (Pacini 1967; Gold 1968). 
Since then, precise, long-term monitoring of radio pulsars has shown the
presence of timing irregularites in most if not all sources.  In isolated
pulsars, these deviations can be grouped into two broad categories: ($i$)
glitches or discontinuous increases in the spin frequency (Radhakishnan \&
Manchester 1969; Reichley \& Downs 1969) and ($ii$) timing noise which can be
described as a more gradual drifting of the {\it measured} torque on the star
(Boynton et al.\ 1972).

In the last few years, a small group of isolated X-ray pulsars has been
identified with timing properties qualitatively similar to, but more extreme
than those of radio pulsars.  This group comprises nine sources, four of which
are known as Soft Gamma Repeaters (SGRs) and five as Anomalous X-ray Pulsars
(AXPs).  Like radio pulsars, these sources are spinning down and show timing
noise, however, at magnitudes on average one hundred times larger.  Interpreted
as magnetic braking, their very rapid spin-down would suggest SGRs/AXPs have
magnetic fields in the $10^{14}-10^{15}$ G range (Thompson \& Duncan 1996;
Kouveliotou et al.\ 1998, 1999). 

Collectively, the SGRs and AXPs possess persistent luminosities of order
$\sim10^{34}-10^{35}$ ergs s$^{-1}$, soft X-ray spectra (as compared to
accreting systems), spin periods within a narrow range (5$-$12 s), rapid
spin-down rates (frequency derivatives $\sim$10$^{-12}$ Hz s$^{-1}$), and
strong timing noise (see Hurley 2000 and Mereghetti 2000 for recent reviews on
SGRs and AXPs, respectively).  The primary difference between AXPs and SGRs is
that so far, only SGRs have been observed to burst.  SGR bursts often occur in
bunches, with active periods lasting on average a few weeks, recurring at time
intervals of years (e.g.\ {G\"o\u{g}\"u\c{s}} et al.\ 2001).  Occasionally,
they emit extremely long and bright bursts, now known as ``giant flares''. 
Only two such flares have been recorded so far: one from SGR~0526$-$66 on March
5, 1979 (Mazets et al.\ 1979) and more recently one from SGR~1900$+$14 on
August 27, 1998 (Hurley et al.\ 1999a; Feroci et al.\ 1999, 2001; Mazets et
al.\ 1999).  In each of these flares, a hard, bright initial spike (peak
luminosity $\sim10^{44}$ ergs s$^{-1}$) was followed by a slowly decaying,
several minute long tail showing coherent pulsations at periods of 8 and 5 s,
respectively.  In fact, the existence of strongly magnetized neutron stars or
`magnetars' (Duncan \& Thompson 1992) in SGRs was first proposed based
primarily upon the first of these giant flares from SGR~0526$-$66 (Duncan \&
Thompson 1992; Paczy\'nski 1992; Thompson \& Duncan 1995).

In the magnetar model, it is the decay of the strong magnetic field which
powers both the persistent and burst emissions (Thompson \& Duncan 1995,
1996).  The short duration SGR bursts are believed to be triggered by
starquakes induced by magnetic stresses in the neutron star crust.  Cracking
the crust leads to a sudden injection of Alfv\'en waves into the magnetosphere,
particle acceleration, and ultimately, the formation of an optically thick pair
plasma which cools and radiates.  The giant flares involve a more profound
restructuring of the crust and magnetic field on a global scale (Thompson \&
Duncan 1995; Woods et al.\ 2001; Ioka 2001).  The persistent X-ray emission
from magnetars is believed to be due to persistent magnetospheric currents
driven by twists in the evolving magnetic field (Thompson \& Duncan 1996),
thermal emission from the stellar surface (e.g.\ \"Ozel 2001) heated by the
decay of the strong field (Thompson \& Duncan 1996), or some combination of the
two.  In any hybrid model, however, the pulse fraction of each spectral
component must be nearly the same in order to account for the weak or absent
energy dependence of the pulse fraction in these sources (\"Ozel, Psaltis \&
Kaspi 2001). The rapid spindown of the neutron star is
believed to be due to magnetic dipole radiation in combination with a wind of
relativistic particles, qualitatively similar to radio pulsars.

Alternatives to the magnetar model for SGRs and AXPs invoke accretion to
explain their persistent X-ray luminosities.  In this scenario, accretion
torques act upon a neutron star with a `standard' field strength ($\sim10^{12}$
G) spinning it down rapidly.  In line with the strict constraints on binary
companions accompanying the SGRs and AXPs (e.g.\ Mereghetti, Israel \& Stella
1998; Wilson et al.\ 1999; Woods et al.\ 2000; Patel et al.\ 2001), the neutron
star is still isolated, yet retains a fallback disk left over from the
supernova explosion that formed the neutron star.  These models naturally
explain the narrow period distribution and can account for the relatively
strong torque variability in general.  However, these models fail to adequately
explain the super-Eddington bursts in the case of SGRs.  Furthermore, for some
AXPs (e.g.\ Hulleman, van Kerkwijk \& Kulkarni 2000; Hulleman et al.\ 2000) and
SGR~0526$-$66 (Kaplan et al.\ 2001), the observed brightness levels and in some
cases upper limits of optical/IR counterparts are well below what is expected
from reprocessed X-ray emission within a disk.  Finally, the detection of
optical pulsations at high rms from 4U~0142$+$61 (Kern \& Martin 2001) are
implausible in the context of this model.

Here, we report on our long-term X-ray observations of SGR~1806$-$20 and
SGR~1900$+$14 using predominantly data from the {\it RXTE} PCA.  We show that
each SGR undergoes an extended ($\gtrsim$ 1 year) interval of accelerated
spin-down where the measured torque on the star increases by a factor $\sim$4. 
We directly compare the burst activity of each source with the long-term torque
variability.  The timing noise is quantified in two ways.  We measure the
timing noise parameter $\Delta$ and also construct torque noise power spectra
for each source.  The timing noise measurements for these SGRs are compared
with the measured noise strengths in known accreting systems and radio
pulsars.  Finally, we discuss implications these results have on the models for
torque variability in magnetars, and what constraints they impose upon SGR
models in general.

\section{SGR~1900$+$14}

Shortly after the discovery of pulsations from SGR~1900$+$14 in 1998 (Hurley et
al.\ 1999b), it was realized that the source did not have a constant spin-down
rate (Kouveliotou et al.\ 1999).  Using archival data, it was determined that
the {\it average} spin-down rate between 1996 September and 1998 June was
$\sim-$2.3 $\times$ 10$^{-12}$ Hz s$^{-1}$ (Marsden et al.\ 1999; Woods et al.\
1999a), however, the local rate in 1996 September (far removed from burst
activity) was significantly higher, at $-$3.1 $\times$ 10$^{-12}$ Hz s$^{-1}$
(Woods et al.\ 1999b).

The SGR became active again in May 1998 and burst for almost one year, during
which a giant flare was observed on 1998 August 27 (Hurley et al.\ 1999a;
Mazets et al.\ 1999; Feroci et al.\ 1999, 2001).  Between 1998 June and the day
following the giant flare, the average spin-down rate doubled.  Due to the
abscence of frequency measurements during this time interval, it was not known
whether this accelerated spin-down was gradual, spanning the gap in
observations, or was instead a sudden jump, perhaps linked with the giant flare
(Woods et al.\ 1999b; Thompson et al.\ 2000).  Recently, Palmer (2001) has
shown that the phase of the pulsations during the tail of the flare does not
match that expected from an extrapolation of the ephemeris determined shortly
following the flare ($\sim$1 day).  Assuming that there is no strong energy
dependence in the pulse profile, this suggests that the star underwent a very
rapid spin-down ($\dot{\nu}$ $\sim-$6 $\times$ 10$^{-10}$ Hz s$^{-1}$) during
the hours following the August 27$^{\rm th}$ flare.  This result provides
independent evidence for a transient particle wind blown off the surface during
the flare (Frail, Kulkarni \& Bloom 1999; Thompson et al.\ 2000) and eliminates
the need for a ``braking'' glitch at the time of the flare.  From 1998 August
28 through 1998 October 8, during a very intensive burst active interval, the
spin-down rate was constant at $-$2.23(1) $\times$ 10$^{-12}$ Hz s$^{-1}$,
which is the least rapid spin-down rate observed to date for this source.  In
the following, we present our analysis of X-ray data starting in 1999 January
and continuing through 2001 January.  During these two years, we measure the
largest spin-down rates yet seen from this SGR.

Our first set of observations was obtained with the {\it RXTE} PCA from 1999
January through 1999 July.  Partial results from these data were reported in
Woods et al.\ (1999b).  Here, we have analyzed the full data set.  The sequence
of {\it RXTE} pointings at SGR~1900$+$14 began with a long ($\sim$50 ks)
observation and was followed by 26 shorter ($\sim$10 ks) observations whose
spacing grew from 0.5 days to $\sim$8 days.  The observations were structured
in this manner in order to enable us to phase connect the data (count cycles)
and thereby track the pulse frequency across the full extent of the data set. 
We successfully connected the early observations covering the first 14 days
beginning on MJD 51181 (1999 January 3).  Even during this short span of data,
we were detecting curvature in the phases (i.e.\ spin down).  The next
observation took place eight days later on MJD 51203 (1999 January 25); the
measured phase was inconsistent with an extrapolation of our model.  We checked
to see whether the inconsistency was due to a change in the pulse profile
during the eight day gap in observations, however, there was no measureable
change.  Unfortunately, we were not able to phase connect the remainder of the
data.

In an attempt to recover the pulse phase ephemeris during this epoch, we
constructed Lomb-Scargle power spectra (Lomb 1976) for each observation.  These
power spectra were converted to frequencies and combined into a grayscale plot
to form a dynamic power spectrum (Figure 1) which clearly shows the spin-down
trend of the SGR.  We used these power spectra to perform a grid search in
frequency and frequency derivative space.  Choosing MJD 51280 as our epoch we
searched frequencies from 0.19357 Hz to 0.19387 Hz and frequency derivatives
from $-$20.0 $\times$ 10$^{-12}$ Hz s$^{-1}$ to $+$6.7 $\times$ 10$^{-12}$ Hz
s$^{-1}$.  We assign the power in each grid point by averaging the
intersections of the various power spectra with the simple model (i.e.\
frequency and frequency derivative).  We show a contour map of the power in
this grid in Figure 2.  The frequency (0.193732 Hz) and frequency derivative
($-$3.6 $\times$ 10$^{-12}$ Hz s$^{-1}$) corresponding to the peak power in
this grid is overplotted in Figure 1 as a dashed line.

\begin{figure}[!htb]
\centerline{
\psfig{file=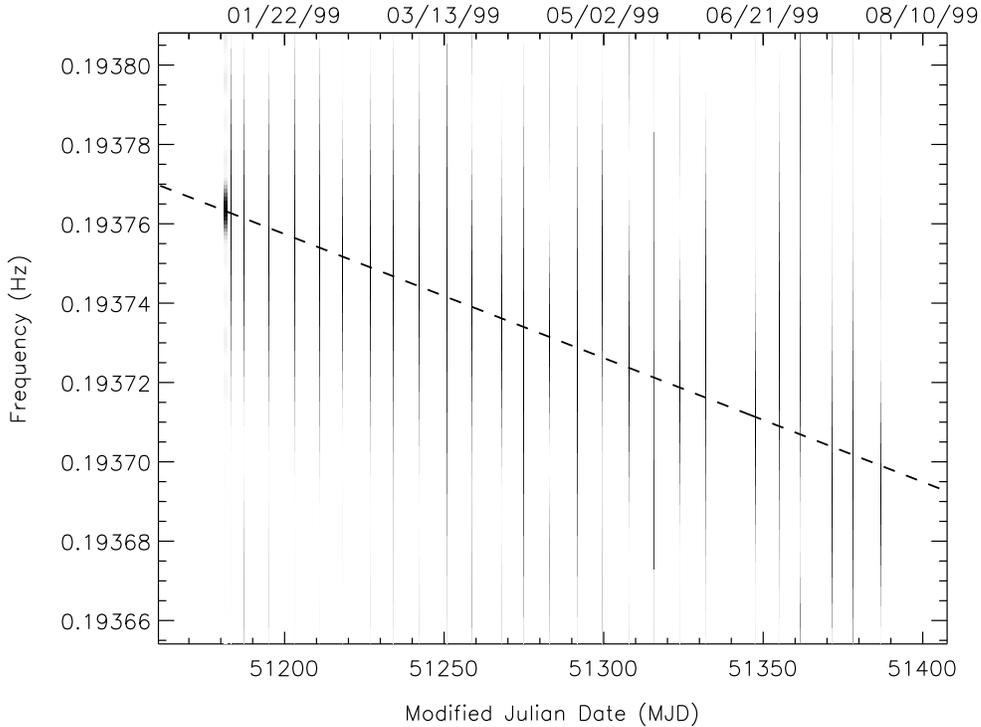,height=4.0in}}
\vspace{-0.05in}

\caption{Dynamic Lomb-Scargle power spectrum of SGR~1900$+$14 during the first
half of 1999.  Darker regions denote higher power.  The dashed line marks the
average ephemeris as determined from a grid search of this power spectrum in
frequency and frequency derivative space (see Fig 2).}

\vspace{11pt}
\end{figure}

\begin{figure}[!htb]
\centerline{
\psfig{file=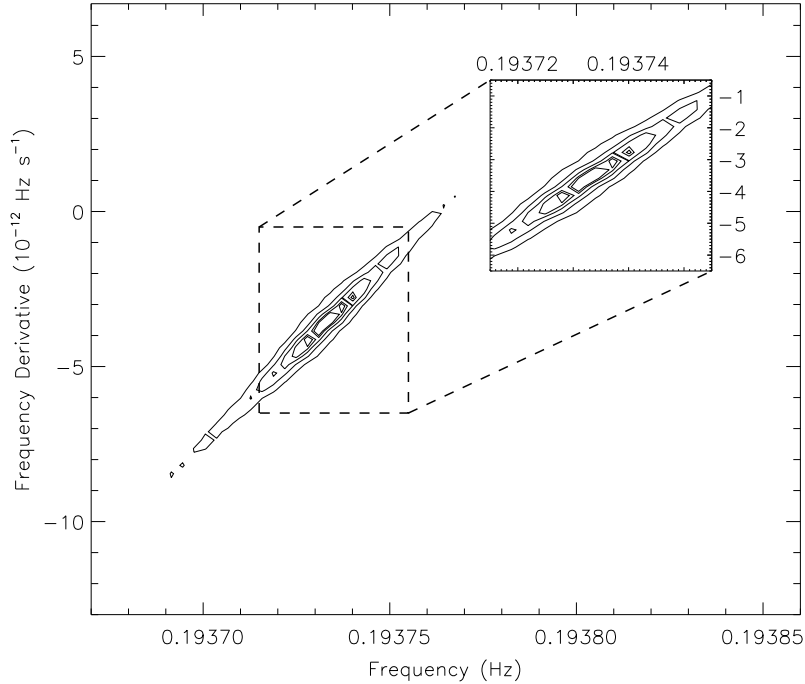,height=4.0in}}
\vspace{-0.15in}

\caption{Contour map of average Lomb power for a grid search to the
SGR~1900$+$14 data shown in Figure 1 (see text for details).}

%\vspace{11pt}
\end{figure}

Using the rough ephemeris found through the grid search described above, we
folded all the data after the gap where we lost track of the phases. 
Unfortunately, we still could not construct a reliable phase connected solution
for these data despite finding no significant changes in the pulse profile. 
Consequently, we are limited to coarse, independent frequency measurements from
the Lomb-Scargle power spectra and a rough ephemeris for the full range of data
provided by the grid search.  The individual frequencies for each observation
are listed in Table 1.

To constrain the manner in which the torque changed early within our observing
campaign, we tried fitting a glitch phase model to the data near the time of
the change allowing for either a positive (ordinary glitch) or negative
(braking glitch) jump in frequency.  We find that neither glitch model provides
a significantly better fit to both the phase and frequency data over a higher
order polynomial with an equal number of free parameters.  Furthermore,
extrapolating the post-glitch ephemerides does not provide an adequate fit to
the data during the following several weeks.  Therefore, we conclude that the
observed drifting in phase of the SGR~1900$+$14 pulsations during 1999 is not
due to a simple shift in frequency and frequency derivative (i.e. a glitch). 
Instead, these data suggest strong timing noise in SGR~1900$+$14 on a short
timescale ($\lesssim$1 week).

The next set of observations of SGR~1900$+$14 took place in 2000 March and
April with the BeppoSAX NFI.  The spectral results have been reported in Woods
et al.\ (2001).  Here, we present the timing results from these data.  For each
observation, we extracted counts from the combined MECS images from a 
4$^{\prime}$ radius circle centered on the source and barycenter corrected the
photon arrival times.  We then constructed a Lomb-Scargle power spectrum across
a narrow range of frequencies (0.192 $-$ 0.194 Hz).  In each observation, we
find a significant peak ($\gtrsim$ 5$\sigma$) that we attribute to the spin
frequency of the neutron star.  We then split each observation into $\sim$12
sections and refined our frequency measurement by fitting a line to the
relative phases of each section.  The spin frequency measurements are listed in
Table 1.  We find that the frequency derivative between 2000 March and April
exceeds the average value observed during both the quiescent interval prior to
burst activity and the most burst active interval (1998 September) by a factor
$\sim$4.

\begin{figure}[!htb]
\centerline{
\psfig{file=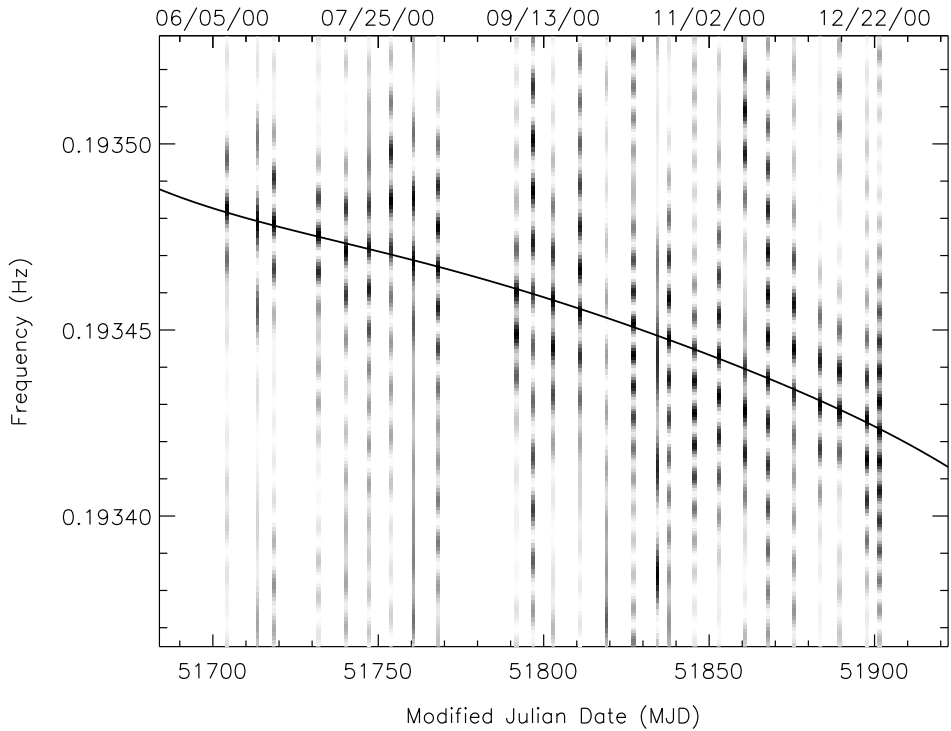,height=4.0in}}
\vspace{-0.05in}

\caption{Dynamic Lomb-Scargle power spectrum of SGR~1900$+$14 during 2000. 
Darker regions denote higher power.  The solid line marks the phase-connected
timing solution to these data.}

\vspace{11pt}
\end{figure}

We began a long-term monitoring campaign of SGR~1900$+$14 with the {\it RXTE}
PCA early in 2000.  These observations continued with only minimal interuption
through 2001 January.  Initially, the schedule was structured as before with a
single observation for each week.  In 2000 June, the observing schedule was
greatly improved by splitting the observations into pairs of $\sim$10 ks
exposure each separated by several spacecraft orbits.  Extending the temporal
baseline of the observations smears out the pulsed signal, but increases the
frequency resolution.  As with the PCA data from 1999, we constructed a dynamic
Lomb-Scargle power spectrum (Figure 3).  The frequency resolution (i.e.\ band
separation in Figure 3) varies with changes in the spacing of each pair of
observations.  This enabled us to better track the pulse frequency through the
observations and to phase connect the entire data set.  The phase residuals are
shown in Figure 4 after subtraction of a second order polynomial ({\it top})
and a sixth order polynomial ({\it bottom}).  The residuals found after
removing the sixth order polynomial are consistent with the measurement
uncertainties ($\chi^2$ = 23.3 for 20 dof). The best fit ephemeris is reported
in Table 2 and overplotted on Figure 3 (solid line).  Within this 210 day
stretch of data the frequency derivative changes by nearly a factor of 2, but
the pulsed flux is consistent with remaining constant.

\begin{figure}[!htb]
\centerline{
\psfig{file=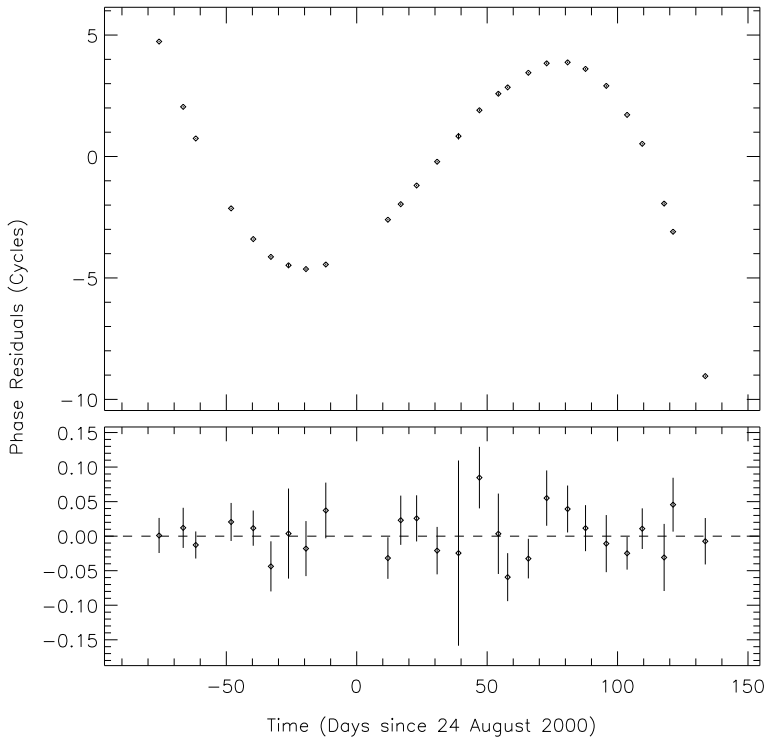,height=3.9in,%
bbllx=58bp,bblly=216bp,bburx=576bp,bbury=672bp,clip=}}
\vspace{-0.10in}

\caption{Phase residuals for SGR~1900$+$14 during year 2000 minus a quadratic
trend ({\it top}) and minus a 6$^{\rm th}$ order polynomial ({\it bottom}).}

\vspace{11pt}
\end{figure}

We have combined the timing results presented here with previous frequency
measurements in order to construct a frequency history of SGR~1900$+$14 (Figure
5).  The plotting symbols mark individual frequency measurements and the solid
lines denote phase-connected timing solutions.  Note that some line segments
are too short to be seen on this scale.  Throughout the 4.3 year frequency
history of SGR~1900$+$14, we find no evidence for episodes of spin-up.  We note
that the overall coverage of observations is fairly sparse, so small amplitude
short-term spin-up episodes cannot be excluded.  The dashed line is the average
spin-down rate measured prior to burst activation of the source in 1998. 
Extrapolation of this early trend in the spin-down clearly shows the large
changes in torque imparted upon SGR~1900$+$14.

\begin{figure}[!p]
\centerline{
\psfig{file=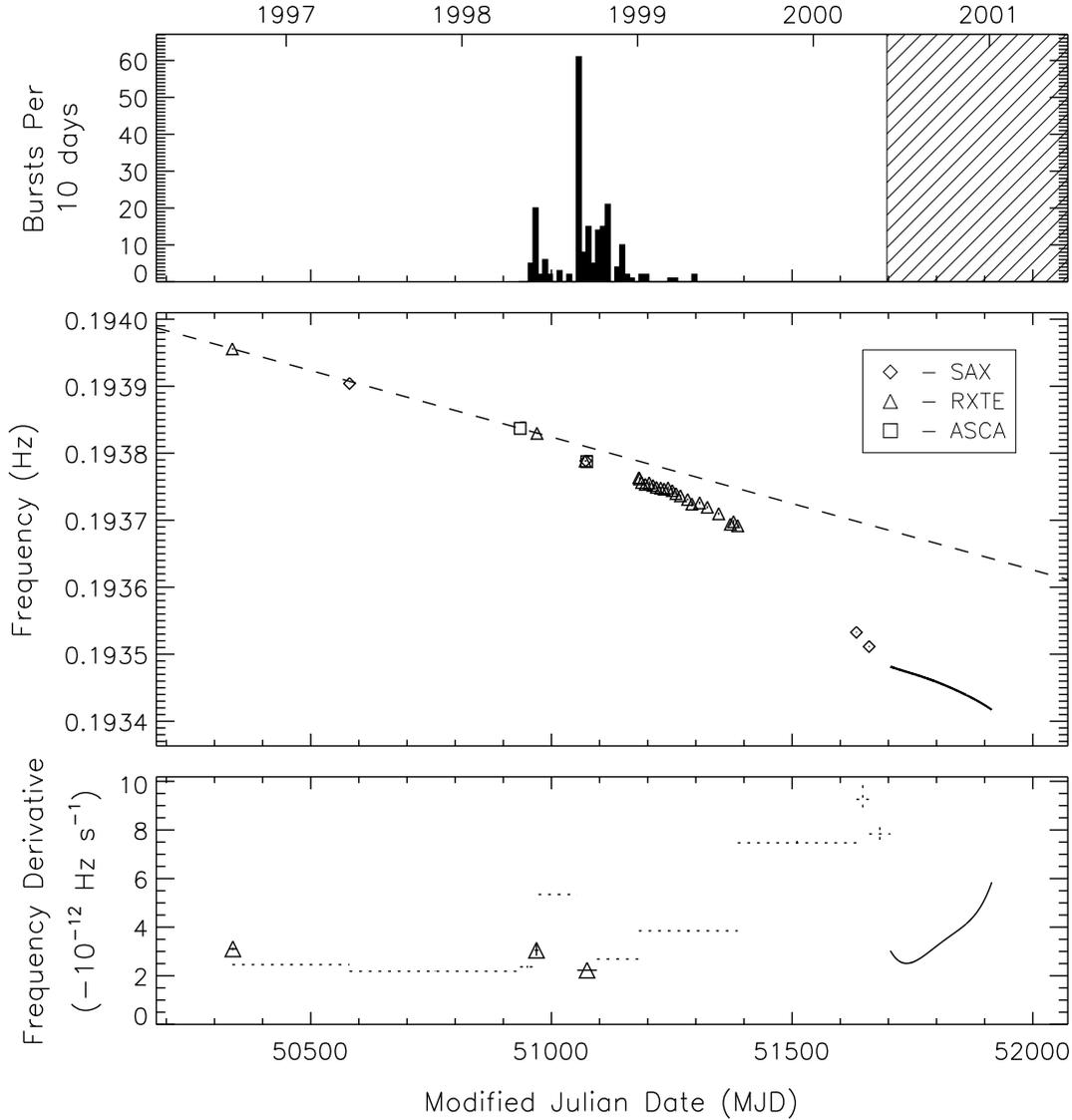,height=6.5in}}
\vspace{-0.15in}

\caption{{\it Top} -- Burst rate history of SGR~1900$+$14 as observed with
BATSE.  The hashed region starts at the end of the {\it CGRO} mission.  {\it
Middle} -- The frequency history of SGR~1900$+$14 covering 4.3 years.  Plotting
symbols mark individual frequency measurements and solid lines denote
phase-connected timing solutions.  The dashed line marks the average spin-down
rate prior to burst activation in 1998.  {\it Bottom} -- The frequency
derivative history over the same timespan.  Dotted lines denote average
frequency derivative levels between widely spaced frequency measurements. 
Solid lines mark phase-coherent timing solutions and triangles mark
instantaneous torque measurements, both using {\it RXTE} PCA data.}

\vspace{11pt}
\end{figure}

Plotted above the frequency history is the burst rate history of SGR~1900$+$14
(Woods et al.\ 1999b) as observed with the Burst and Transient Source
Experiment (BATSE) onboard the {\it Compton Gamma-ray Observatory (CGRO)}.  The
hashed region starting in 2000 June marks the end of the source monitoring with
BATSE after the demise of {\it CGRO}.  It is clear from this figure that the
periods of enhanced torque do not correlate directly with the burst activity,
confirming earlier results (Woods et al.\ 1999b).  However, one cannot rule out
a causal relationship between the two with a delay of several months between
the burst activity and the subsequent torque variability.  Unfortunately, the
limited coverage of the persistent X-ray source during the last 20 years does
not allow concrete conlusions on this relationship (see also \S5).

\section{SGR~1806$-$20}

\begin{figure}[!b]
\centerline{
\psfig{file=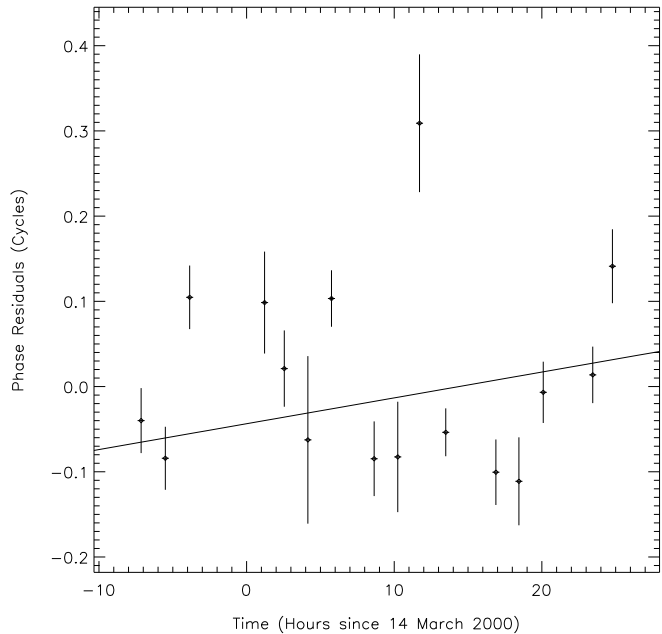,height=3.7in}}
\vspace{-0.15in}

\caption{Phases for SGR~1806$-$20 during a long observation in March 2000.  The
solid line denotes a least-squares fit to the data.}

\vspace{11pt}
\end{figure}

From 1993 through 1998, there were only three X-ray observations of
SGR~1806$-$20.  Using these observations, Kouveliotou et al.\ (1998) determined
that the SGR had an average spin-down rate of $-$1.5 $\times$ 10$^{-12}$ Hz
s$^{-1}$ over this time interval.  Early in 1999, we initiated a sequence of
PCA observations of this SGR which lasted for 178 days.  The data were
successfully phase connected and although the long-term frequency derivative
did not change by very much, significant short-term deviations (timescale of
months) were found (Woods et al.\ 2000).  To model these deviations, we fit a
constant spin-down plus an orbit to all available phases and frequencies
between 1993 and 1999; the fit was statistically acceptable.  Given the
sparseness of the data, however, it was not possible to determine whether the
observed torque variations were due to orbital effects or strong timing noise. 
We have recently completed the analysis of new {\it RXTE} PCA observations
during year 2000 (January through November) to address this question.

Early in January 2000, SGR~1806$-$20 entered a moderately burst active state
that led us to initiate densely sampled ToO observations with the PCA.  The
initial observation contained two notable surprises.  First, the measured
frequency (Table 3) was substantially smaller than expected based upon an
extrapolation of the historical frequency history, effectively ruling out
previously acceptable orbital solutions (Woods et al.\ 2000).  In addition, the
phase residuals (Figure 6) were unusually large ($\chi^2$ = 73.1 for 14 dof)
for such a relatively short stretch of data (1.3 days).  The pulse profile
appears sinusoidal during these observations showing no gross changes
throughout.  This is the shortest timescale ($\sim$hours) over which we have
observed timing noise in SGR~1806$-$20 thus far.  We note, however, that
SGR~1900$+$14 displayed timing noise on an even shorter timescale
($\sim$minutes) during {\it Chandra} observations following a recent burst
activation of this source in 2001 (Kouveliotou et al.\ 2001).

\begin{figure}[!hb]
\centerline{
\psfig{file=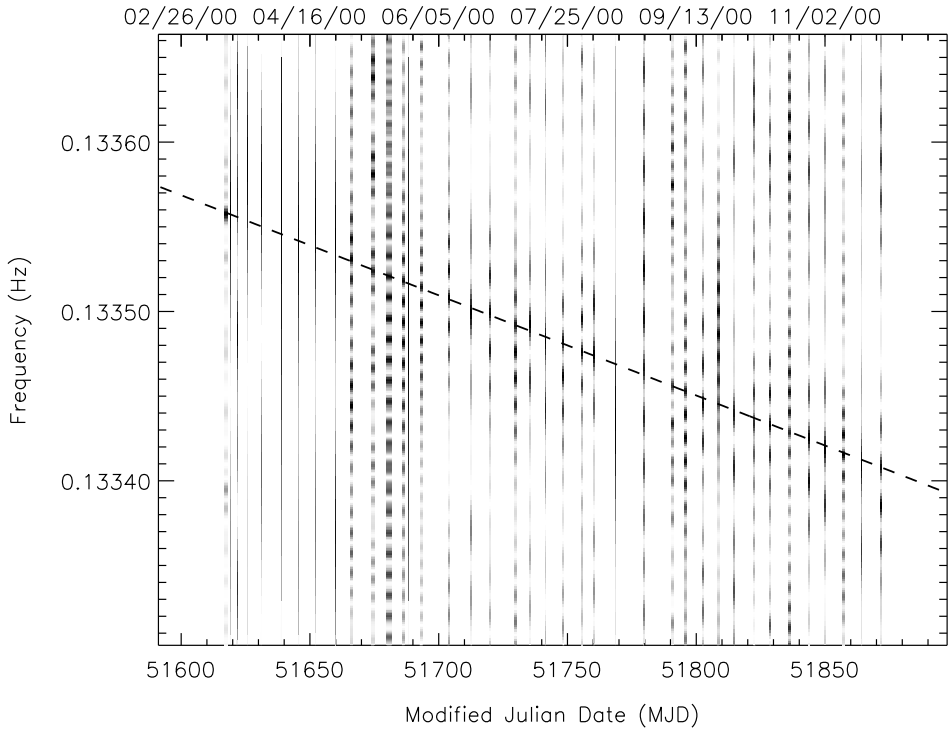,height=4.0in}}
\vspace{-0.05in}

\caption{Dynamic Lomb-Scargle power spectrum of SGR~1806$-$20 during the year
2000.  Darker regions denote higher power.  The dashed line marks the average
ephemeris as determined from a grid search of this power spectrum in frequency
and frequency derivative space (see Figure 8).}

\vspace{11pt}
\end{figure}

After this initial observation, we began brief ($\sim$5 ks) weekly observations
of the source.  The torque variations continued, negating the possibility of
achieving long baseline phase coherent solutions.  In 2000 May, the weekly
observation schedule was restructured employing the same strategy as was done
for SGR~1900$+$14.  For SGR~1806$-$20, exposure time of each pointing was
$\sim$5 ks with a total of 10 ks per pair per week.  Observations of
SGR~1806$-$20 continued until the source was no longer observable by {\it RXTE}
due to Sun angle constraints in November.  We next constructed a dynamic
Lomb-Scargle power spectrum using these data (Figure 7) and performed a grid
search in frequency (0.13340 to 0.13355 Hz) and frequency derivative ($-$20.0
to $+$6.7 $\times$ 10$^{-12}$ Hz s$^{-1}$) space.  The Lomb-Scargle power map
(Figure 8) reaches a maximum at 0.1334800 Hz and $-$6.85 $\times$ 10$^{-12}$ Hz
s$^{-1}$ for the frequency and frequency derivative, respectively. The epoch
chosen for the grid search was MJD 51750.  This best fit average ephemeris is
overplotted in Figure 7.

\begin{figure}[!htb]
\centerline{
\psfig{file=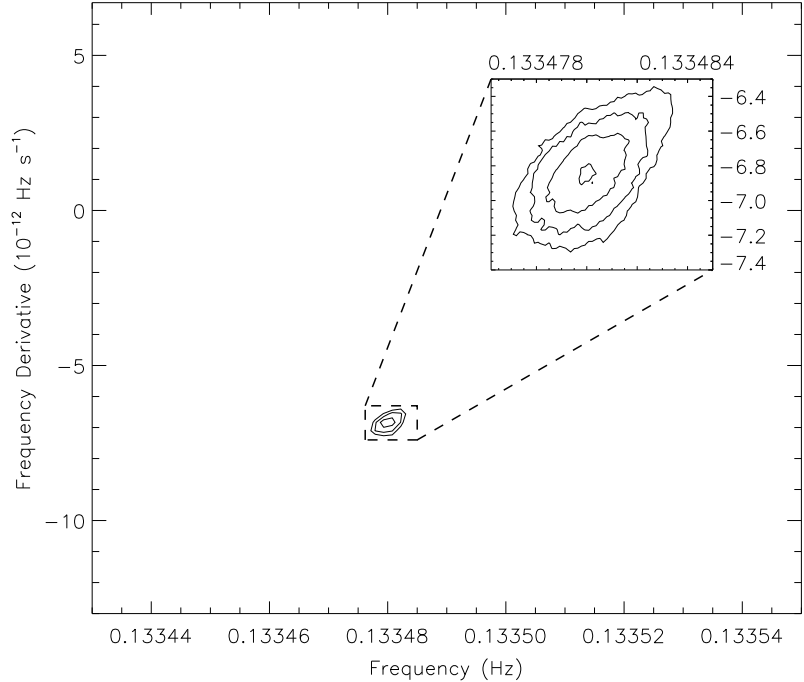,height=4.0in}}
\vspace{-0.15in}

\caption{Contour map of average Lomb power for a grid search to the
SGR~1806$-$20 data shown in Figure 7 (see text for details).}

\vspace{11pt}
\end{figure}

We find that the spin-down rate during the year 2000 is a factor $\sim$4 larger
than the average rate between 1993 and 1999.  Unlike SGR~1900$+$14, we were not
able to phase connect across the full extent of these data.  However, we were
able to connect a 55 day subset between MJD 51666 and 51720 (Table 4).  For a
77 day subset between MJD 51796 and 51872, we attained a formally acceptable
fit to the phases and frequencies.  However, we are not fully confident that
there are no cycle count ambiguities in this span of data.  Coarse frequency
measurements for the remaining data are reported in Table 3.

\begin{figure}[!p]
\centerline{
\psfig{file=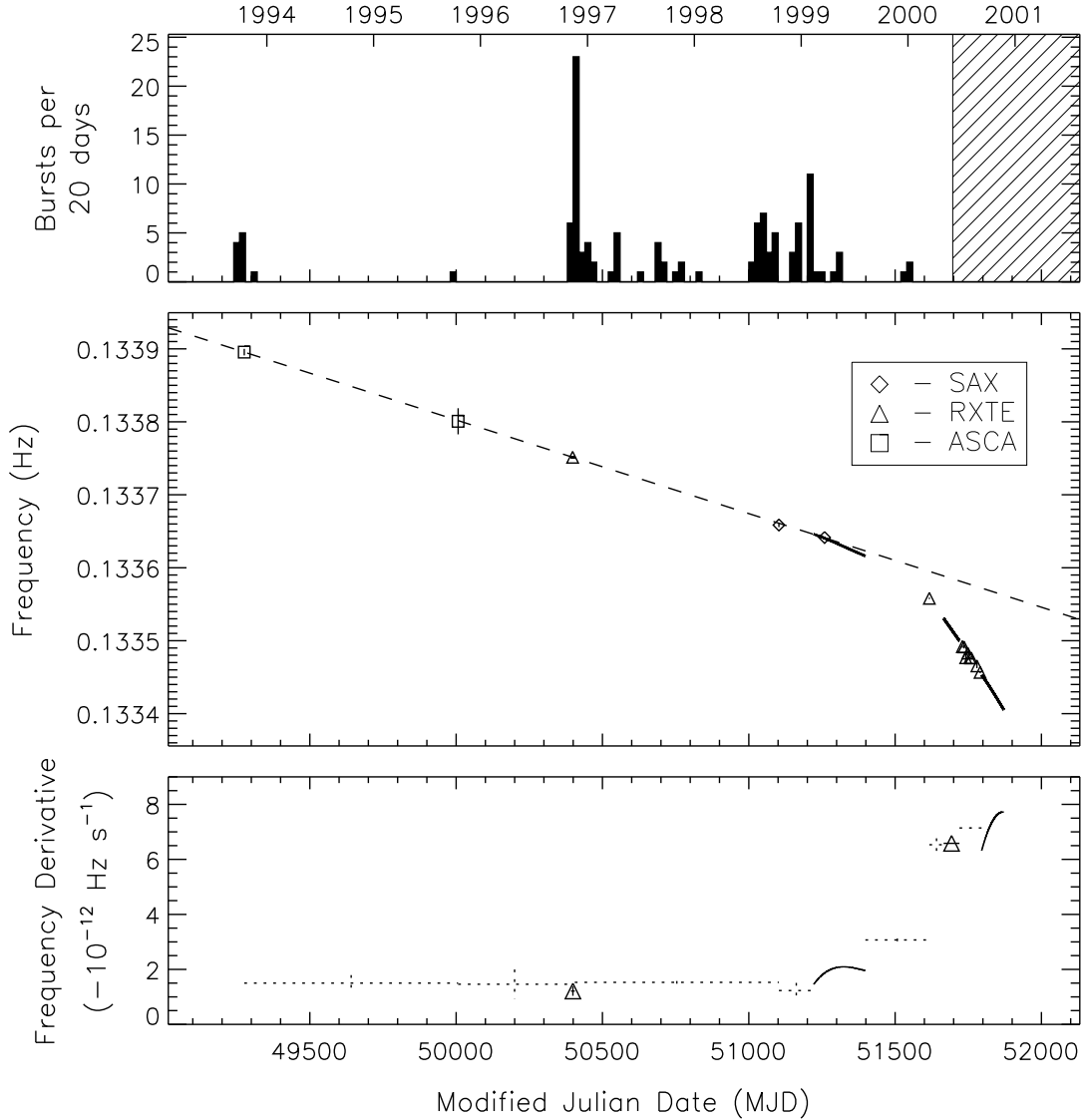,height=6.5in}}
\vspace{-0.15in}

\caption{{\it Top} -- Burst rate history of SGR~1806$-$20 as observed with
BATSE.  The hashed region starts at the end of the {\it CGRO} mission.  {\it
Middle} -- The frequency history of SGR~1806$-$20 covering 7.1 years.  Plotting
symbols mark individual frequency measurements and solid lines denote
phase-connected timing solutions.  The dashed line marks the average spin-down
rate prior to burst activation in 1998.  {\it Bottom} -- The frequency
derivative history over the same timespan.  Dotted lines denote average
frequency derivative levels between widely spaced frequency measurements. 
Solid lines mark phase-coherent timing solutions and triangles mark
instantaneous torque measurements, both using {\it RXTE} PCA data.}

\vspace{11pt}
\end{figure}

The current frequency history of SGR~1806$-$20 from year 1993 through 2000 is
shown in Figure 9.  As with Figure 5, the symbols mark individual frequencies
and solid lines denote phase-connected timing solutions.  There is no evidence
for spin-up in SGR~1806$-$20 during the last 7.1 years.  Although, as with
SGR~1900$+$14, we note the caveat that small amplitude short-term spin-up
episodes cannot be excluded due to the sparseness of the observations.  The
dashed line represents the average spin-down rate during the $\sim$3 years
prior to the 1996 burst activation of the source.  At some point during late
1999, the average spin-down rate of SGR~1806$-$20 quadrupled and has not yet
returned to the pre-1999 rate.

The burst rate history of SGR~1806$-$20 as observed with BATSE is shown in the
top panel of Figure 9.  BATSE monitored the burst activity through the large
increase in average torque in late 1999, but detected nothing out of the
ordinary.  As with SGR~1900$+$14, this interval of accelerated spin-down does
not correspond to a period of strong burst activity in the SGR, but rather
follows a moderate one.  There is no apparent connection between the burst
activity in SGR~1806$-$20 and the measured torque enhancement.

\section{Timing Noise}

Strong timing noise is present in both SGR~1806$-$20 and SGR~1900$+$14.  There
are numerous ways with which to characterize timing noise in pulsars.  Here, we
have chosen to quantify the timing noise levels in these SGRs using two
separate methods in order to make comparisons to samples of both accreting
systems and canonical radio pulsars.  First, for each SGR, we obtain an
instantaneous estimate of the timing noise parameter $\Delta$ (Arzoumanian et
al.\ 1994).  Next, using the complete frequency history, we generate torque
noise power density spectra for each source (Deeter \& Boynton 1982).

\subsection{$\Delta$ Parameter}

The timing noise parameter, $\Delta$, is defined as $\Delta_t = {\rm
log}(\frac{|\ddot{\nu}| t^3}{6\nu})$.  This characterization of timing noise
was introduced by Arzoumanian et al.\ (1994) for radio pulsars and later
applied to a subset of the AXP population by Heyl \& Hernquist (1999).  For the
radio pulsar data analyzed by Arzoumanian et al., the average timespan of the
observations for each source in the sample was a few years or $\sim$10$^8$ s. 
Therefore, $\Delta_8$ was measured for each radio pulsar.  Earlier, we
calculated $\Delta_8$ for SGR~1806$-$20 using a 178 day interval ($\sim$10$^7$
s) of phase connected data where we could directly measure $\ddot{\nu}$ (Woods
et al.\ 2000).

For SGR~1900$+$14, the longest stretch of phase connected data is 210 days in
the year 2000.  As was done for SGR~1806$-$20, we measured the average
$\ddot{\nu}$ over the full extent of the data and calculated $\Delta_8$.  We
measure $\Delta_8 =$ 5.3, very similar to that found for SGR~1806$-$20
($\Delta_8 =$ 4.8 [Woods et al.\ 2000]).

The $\Delta_8$ values measured for these SGRs by far exceed any of those in the
large sample of radio pulsars.  Arzoumanian et al.\ (1994) noted that the
strength of the timing noise parameter correlates with the period derivative. 
The two parameters scale approximately according to $\Delta_8 = 6.6 + 0.6~{\rm
log}(\dot{P})$, albeit with considerable scatter about this trend.  Having
average period derivative values $\dot{P} \sim 10^{-10}$ s s$^{-1}$, the
observed $\Delta_8$ values of these SGRs are considerably higher than the trend
(difference $\sim$4.4), whereas 97\% of the measured radio pulsars in the
Arzoumanian et al.\ sample fall within $\pm$1.5 in $\Delta_8$ about the trend. 
Having estimated $\Delta_8$ for some of the quieter accreting sources (Woods et
al.\ 2000) using extended sequences of phase connected data (Chakrabarty 1996),
we find that these two SGRs are ``quieter'' than the majority of accreting
sources, yet within the range of observed timing noise strengths.  Using the
$\Delta_8$ parameterization, we conclude that the timing noise in these SGRs is
more in line with the accreting sources.  However, $\Delta_8$ is a relatively
crude parameterization of timing noise.  Given the longer temporal baseline now
available to us for each SGR, we can construct meaningful torque noise power
spectra which will allow us to make more detailed comparisons with accreting
systems and radio pulsars.

\subsection{Torque Noise Power Spectra}

For each SGR, we constructed torque noise power density spectra following
Deeter (1984).  This technique was developed to construct power spectra for
unevenly sampled data sets with non-uniform errors and is therefore very
suitable for application to our SGR data.  The method utilizes polynomials
instead of sinusoids to estimate the power of torque fluctuations at a given
analysis frequency.  For the lower analysis frequencies ($\lesssim10^{-7.2}$
Hz), we used the spin frequency histories of each SGR and fit cubic polynomial
estimators.  The higher analysis frequencies ($\gtrsim10^{-7.2}$ Hz) were
probed using the long stretches of phase-connected data with quartic
estimators.  For a more detailed description of the methodology, see \S5.4 of
Bildsten et al.\ (1997).

The SGR torque noise power spectra are extremely red and are consistent with a
steep power-law.  We fit these power spectra to a power-law with ${\rm
log}(S_{\dot{\nu}}) = \alpha_{7.5} + \beta~{\rm log}(f_s)$, where
$S_{\dot{\nu}}$ is the power density of the frequency derivative, $f_s$ is the
analysis frequency, $\alpha_{7.5}$ is the normalization at 10$^{-7.5}$ Hz, and
$\beta$ is the power-law index.  However, a simple least squares fit to the
power estimates is inadequate to accurately measure these quantities and their
errors.  The principle difficulties in fitting the power spectra is that each
point is correlated with the others, the mid-response frequency calculated for
each power measurement is dependent upon the power-law index, the errors on the
power levels are only approximations within this methodology, and these errors
are highly asymmetric. 

To accurately parameterize the power spectra, we set up Monte Carlo simulations
for each SGR data set.  Only power estimates significantly above the
measurement noise level ($\lesssim$10$^{-6.9}$ Hz) were utilized for this
simulation.  Phases and spin frequencies at each epoch were simulated according
to an assumed power-law torque noise power density spectrum.  One thousand
realizations were constructed for a 6x6 grid of power-law normalizations and
slopes with ranges which encompassed the measured values from the raw power
spectra.  Comparing the measured powers from the simulated data with the values
input into the simulation, we found that there is a slight bias toward lower
powers at the lower analysis frequencies for the SGR data (i.e.\ the true power
is underestimated).  We searched the grid for the average measured power-law
parameters from the simulation which most closely agreed with those measured
from the SGR data.  We used the results of this simulation to remove the bias
from the power measurements and to estimate the errors on the individual powers
and the measured power-law normalization and slope.

Power-law fits to the corrected SGR data yield normalizations ($\alpha_{7.5}$)
of $-$16.36~$\pm$~0.22 and $-$17.34~$\pm$~0.40 log(Hz$^2$ s$^{-2}$ Hz$^{-1}$)
and power-law indices ($\beta$) of $-$3.7~$\pm$~0.6 and $-$3.6~$\pm$~0.7 for
SGR~1900$+$14 and SGR~1806$-$20, respectively.

\begin{figure}[!htb]
\centerline{
\psfig{file=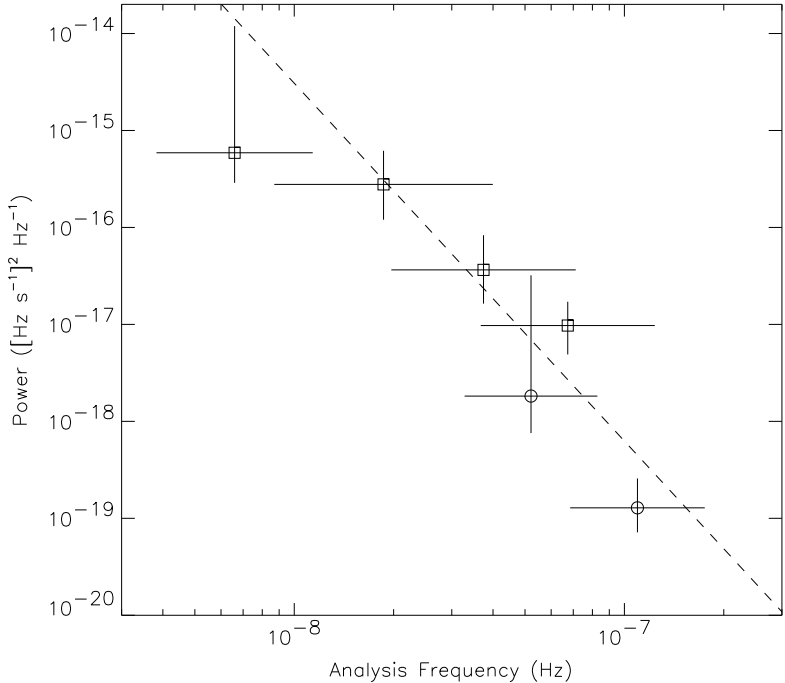,height=4.0in}}
\vspace{-0.15in}

\caption{Torque noise power density spectrum of SGR~1900$+$14.  The squares and
circles denote measurements made with the frequency and phase-connected data,
respectively.  The data shown here have been corrected for biases introduced by
the analysis method (see text for details).  The errors on the power estimates
denote the 68\% confidence intervals which were derived from the Monte Carlo
simulation.  The horizontal bars give the log frequency rms of the estimator
response.  The power levels expected from measurement noise are much lower than
the power measured from the source variability. The dashed line denotes the
best power-law fit to the data.}

\vspace{11pt}
\end{figure}

\begin{figure}[!htb]
\centerline{
\psfig{file=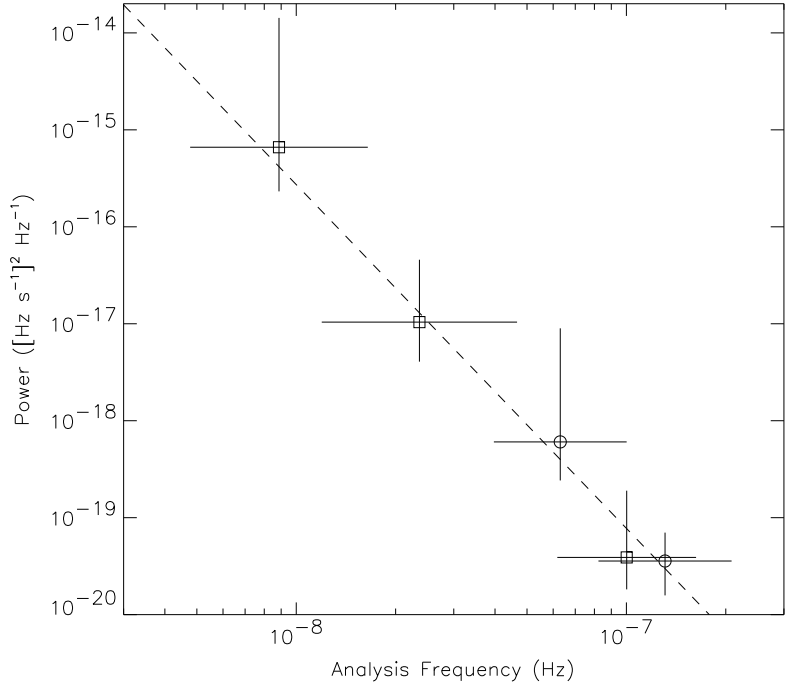,height=4.0in}}
\vspace{-0.15in}

\caption{Torque noise power density spectrum of SGR~1806$-$20.  The squares and
circles denote measurements made with the frequency and phase-connected data,
respectively.  The data shown here have been corrected for biases introduced by
the analysis method (see text for details).  The errors on the power estimates
denote the 68\% confidence intervals which were derived from the Monte Carlo
simulation.  The horizontal bars give the log frequency rms of the estimator
response.  The power levels expected from measurement noise are much lower than
the power measured from the source variability. The dashed line denotes the
best power-law fit to the data.}

\vspace{11pt}
\end{figure}

We next compared the SGR power spectra to those of known accreting sources. 
Bildsten et al.\ (1997) noted that, in general, the wind-fed pulsars had flat
power spectra and disk-fed pulsars obeyed a $f^{-1}$ power-law.  None of the
measured power spectra in their sample are steeper than $\sim f^{-1.2}$.  We
find that the average level of the SGR torque noise power spectra are within
the range of the average power for accreting systems, consistent with the
$\Delta_8$ parameterization.  {\it However, the spectral slopes of the SGRs are
much steeper than any accreting pulsar spectrum.}

Although far more phase-connected timing data are available for radio pulsars,
the sample of published torque noise power spectra is very limited.  From the
available data, it is clear that the power spectra of radio pulsars are much
more diverse than accreting pulsar power spectra.  The power-law indices vary
from somewhat positive values (i.e.\ blue spectra) to very steep negative
values in some cases.  For example, the Vela pulsar has a very red power
spectrum, similar to the SGR power spectra.  From Figure 8 of Alpar et al.\
(1986), we estimate $\alpha_{7.5} \approx -$20.6 and $\beta \approx -$2.7 for
the Vela pulsar.  Note that the power density spectrum normalization and units
in Alpar et al.\ differ from our normalization and units by a factor $\frac {2}
{(2\pi)^2}$.  The factor 2 comes from our excluding power from negative
frequencies and the $(2\pi)^2$ follows from the units conversion (i.e.\ rad
s$^{-1}$ to Hz).  A complete, systematic analysis of radio pulsar data is
required, however, before one can determine where the SGR torque noise power
density spectra fit relative to the radio pulsar population as a whole.

\section{Discussion}

We have shown that the spin-down behaviors of SGR~1900$+$14 and SGR~1806$-$20
are quite erratic, and yet extremely similar to one another.  The time scale of
the torque variations in these SGRs varies from minutes to years and the torque
noise power density spectra of these SGRs are consistent with one another in
both normalization and slope.  The observed range in torque imparted upon these
SGRs varies by up to a factor of $\sim$4 in each case despite gross differences
in the burst activity of the two sources over the same time period. 

There is no evidence for a direct connection between the burst activity and
torque enhancements in these sources.  For example, the most intense burst
active interval for SGR~1900$+$14 was 1998 August $-$ September, during which
the spin-down rate was the lowest measured thus far.  Within the first half of
2000, the spin-down rate of SGR~1900$+$14 was at its highest (more than four
times the rate in the fall of 1998), yet there was no detectable burst emission
during this entire interval.  Similar examples, although less extreme, are
present in the data for SGR~1806$-$20.

Here, we explore the hypothesis that the burst activity and torque variability
may still be related, albeit with a substantial time delay between the two. 
The strongest torque variability observed in these SGRs thus far has been found
{\it following} substantial burst activity.  In the case of SGR~1900$+$14, the
delay would have to be several months; for SGR~1806$-$20, years.  While the
spin frequency of SGR~1900$+$14 has been monitored (1996 $-$ 2001), the source
underwent relatively short intervals of concentrated burst activity including a
giant flare (e.g.\ Hurley et al.\ 1999a), hard burst emission (Woods et al.\
1999c), anomalously long, smooth bursts (Ibrahim et al.\ 2001), and
multi-episodic bursts (e.g.\ Hurley et al.\ 1999c).  For SGR~1806$-$20, the
burst activity between 1996 and 2000 was more persistent and scattered in
time.  In spite of the large differences in burst active states between the two
sources, the relative change in their spin-down rates were nearly the same. 
Given that the delays between burst activity and torque changes are so
different between these sources and there exists no scaling of the relative
change in their spin-down rates for varying degrees of burst activity, we
conclude that there is no causal relationship between the burst activity and
the long-term torque variability.

However, when one considers the AXP and SGR populations as a whole, the burst
activity and torque variability may be linked indirectly through some other
parameter (e.g.\ magnetic field strength).  For the SGRs and AXPs, there
appears to be a trend of increasing noise strength with spin-down rate (Woods
et al.\ 2000; Gavriil \& Kaspi 2002), analogous to the correlation observed in
radio pulsars (Cordes \& Helfand 1980; Arzoumanian et al.\ 1994).  Gavriil \&
Kaspi (2002) suggest that this trend may represent a common bond between all
three populations where the magnitude of the timing noise scales with perhaps
magnetic field strength.  Only sources with the strongest timing noise and
largest spin-down rates (i.e.\ magnetic fields), namely the SGRs, have been
observed to burst.  Therefore, a timing noise/spin-down rate threshold in this
proposed continuum may exist, above which the source will be burst active.  The
proximity of 1E~1048.1$-$5937 to the SGRs in both spin-down rate and timing
noise has led Kaspi et al.\ (2001) to propose that this source as the most
likely candidate to undergo an `SGR-like' outburst in the future.

Our monitoring of the spin behavior of these two SGRs both in burst active
states as well as during quiescence enables us to constrain physical mechanisms
proposed to explain, within the context of the magnetar model, the deviations
from constant spin-down observed in SGRs and AXPs.  It has been argued that a
transient outward surface flux of Alfv\'en waves and particles due presumably
to crust fractures associated with bursts could impart an appreciable change in
torque on a magnetar (Thompson \& Blaes 1998; Harding, Contopoulos \& Kazanas
1999; Thompson et al.\ 2000).  In fact, such an effect was likely detected
(Woods et al.\ 1999b; Palmer 2001) following the extremely luminous flare of 27
August 1998 from SGR~1900$+$14.  However, the more common, shorter bursts have
shown no evidence for associated transient torque changes.  The difference in
energy between the giant flare and the recurrent bursts is several orders of
magnitude ($\gtrsim$3) and the predicted energy dependence of the stellar
angular momentum change goes as $E$ (Thompson et al.\ 2000).  Considering the
energy difference and the short timescale over which the transient 27 August
torque change acted (Palmer 2001), it is not surprising that the other burst
activity from SGR~1900$+$14 and all observed activity from SGR~1806$-$20 have
had no measurable effect on the spin-down of their respective SGRs.

The abscence of a direct correlation between burst activity and torque
enhancements has strong implications for the underlying physics behind each
phenomenon.  The magnetar model postulates that the bursting activity in SGRs
is a result of fracturing of the outer crust of a highly magnetized neutron
star (Thompson \& Duncan 1995).  Furthermore, the majority of models proposed
to explain the torque variability in magnetars invoke crustal motion and/or
low-level seismic activity (e.g.\ Thompson et al.\ 2000).  Since we have found
no direct correlation between the burst activity and torque variability, we
conclude that either ($i$) the seismic activities leading to each observable
are decoupled from one another, or ($ii$) at least one of these phenomena is
{\it not} related to seismic activity.

Precession models, both radiative (Melatos 1999) and free (Thompson et al.\
2000), were proposed previously as possible explanations for the observed
variability in spin-down in SGRs and AXPs.  These models predict periodic
variations in the spin-down on timescales of several months to several years. 
It is clear from the present data that periodic torque variations on short
timescales (i.e.\ less than a few years) can be excluded.  Given the relatively
short baseline of the spin frequency histories for these SGRs, one cannot rule
out precessional periods longer than a few years, although, the strong,
short-term variations must still be explained through some other process.

\section{Conclusions}

Using predominantly {\it RXTE} PCA observations, we have constructed long-term
frequency histories for SGR~1900$+$14 and SGR~1806$-$20. Each SGR exhibits
large excursions from constant spin-down although neither has shown evidence
for spin-up.  The torque variability does not correlate with the burst activity
in either source, contrary to predictions from some models.  The absence of a
correlation between burst activity and torque variability places significant
constraints on the physical mechanisms behind each phenomenon in the context of
the magnetar model.  Specifically, either the seismic activities leading to the
burst activity and torque variability are physically decoupled from one another
or at least one of these phenomena are not related to seismic activity.

Using phase-connected timing solutions and the global frequency histories, we
construct torque noise power density spectra for each SGR.  The torque noise
power spectra are consistent with one another, but do not match the power
spectra of accreting systems or standard radio pulsars.  The SGR torque noise
spectra are much steeper than power spectra of accreting systems and the
frequency averaged power levels are much higher than the spectra of radio
pulsars.  Construction and modeling of torque noise power density spectra for
AXPs and a large collection of radio pulsars would provide a useful
quantitative means for comparison of timing noise behavior between these three
classes of neutron stars.

\acknowledgments{\noindent {\it Acknowledgements} -- We greatly appreciate the
help we received from the RXTE team, particularly Evan Smith and the SOF for
scheduling these extensive observations.  We also thank the RXTE/SDC, the
SAX/SDC, and HEASARC for pre-processing the RXTE/PCA and BeppoSAX data.  This
work was funded primarily through a Long Term Space Astrophysics program (NAG
5-9350) for both PMW and CK.  MHF and EG acknowledge support from the
cooperative agreement NCC 8-200.  PMW, CK, and EG appreciate useful discussions
at the ITP funded by NSF grant PHY99-07949. }

\begin{table}[!hp]
%\vspace{1.3in}
\begin{center}
%\tighten
%\footnotesize
\caption{Pulse frequencies for SGR~1900$+$14.
\label{tbl:1}}
\vspace{11pt}
%\small
\tablewidth{6.0truein}

\begin{tabular}{ccl}
\hline
\hline

Date          &
MJD           &
Frequency     \\

mm/dd/yy      &
TDB           &
(Hz)          \\

\hline

    01/03/99     &  51181.590     &  0.19376364(23)   \\
    01/05/99     &  51183.210     &  0.1937596(28)   \\
    01/09/99     &  51187.274     &  0.193760(4)   \\
    01/17/99     &  51195.073     &  0.193753(4)   \\
    01/25/99     &  51203.069     &  0.193762(6)   \\
    02/01/99     &  51210.940     &  0.193759(5)   \\
    02/09/99     &  51218.158     &  0.1937475(28)   \\
    02/17/99     &  51226.992     &  0.193750(4)   \\
    02/25/99     &  51234.055     &  0.193744(10)   \\
    03/05/99     &  51242.178     &  0.193744(7)   \\
    03/13/99     &  51250.882     &  0.193741(12)   \\
    03/21/99     &  51258.782     &  0.193733(9)   \\
    03/31/99     &  51268.009     &  0.193740(5)   \\
    04/06/99     &  51274.870     &  0.193735(9)   \\
    04/14/99     &  51282.969     &  0.1937307(27)   \\
    04/23/99     &  51291.794     &  0.193726(9)   \\
    05/01/99     &  51299.691     &  0.193759(13)   \\
    05/09/99     &  51307.883     &  0.193723(7)   \\
    05/17/99     &  51315.882     &  0.19372(7)   \\
    05/25/99     &  51323.848     &  0.193724(5)   \\
    06/02/99     &  51331.865     &  0.193720(17)   \\
    06/18/99     &  51347.427     &  0.193710(9)   \\
    06/25/99     &  51354.920     &  0.193733(10)   \\
    07/02/99     &  51361.640     &  0.193730(11)   \\
    07/12/99     &  51371.693     &  0.193699(13)   \\
    07/19/99     &  51378.172     &  0.193697(14)   \\
    07/27/99     &  51386.918     &  0.193690(6)   \\
    03/30/00\tablenotemark{a}     &  51633.500        &  0.1935327(9)   \\
    04/26/00\tablenotemark{a}     &  51660.000        &  0.1935115(10)   \\
    
\hline

\end{tabular}

\begin{flushleft}

{\small 
a -- {\it BeppoSAX} observation: all others use {\it RXTE} PCA
data}

\end{flushleft}

\end{center}
\end{table}

\begin{table}[!hp]
%\vspace{1.3in}
\begin{center}
\tighten
%\footnotesize
\caption{Year 2000 pulse ephemeris for SGR~1900$+$14 from {\it RXTE} PCA 
observations.
\label{tbl:2}}
%\tablewidth{0pt}
\vspace{11pt}
%\small

\begin{tabular}{lc}
\hline
\hline

Parameter           &
Ephemeris 2000      \\

\hline

Date Range                             &  2000 June 09 $-$ 2001 January 04  \\

MJD Range                              &  51704.2 $-$ 51914.9   \\ 

Epoch (MJD)                            &  51780.0   \\ 

$\chi^2$/dof                           &  23.3/20   \\ 

$\nu$ (Hz)                             &  0.193464095(8)    \\ 

$\dot{\nu}$ (10$^{-12}$ Hz s$^{-1}$)   &  -2.913(3)    \\ 

$\ddot{\nu}$ (10$^{-19}$ Hz s$^{-2}$)  &  -1.72(3)    \\ 

$\nu^{(3)}$ (10$^{-26}$ Hz s$^{-3}$)   &  -1.21(8)    \\ 

$\nu^{(4)}$ (10$^{-32}$ Hz s$^{-4}$)   &  1.33(9)    \\ 

$\nu^{(5)}$ (10$^{-39}$ Hz s$^{-5}$)   &  -4.7(4)    \\

\hline

\end{tabular}

\end{center}
\end{table}

\begin{table}[!hp]
%\vspace{1.3in}
\begin{center}
\tighten
%\footnotesize
\caption{Pulse frequencies for SGR~1806$-$20 from {\it RXTE} PCA
observations.
\label{tbl:3}}
%\tablewidth{0pt}
\vspace{11pt}
%\small

\begin{tabular}{ccl}
\hline
\hline

Date             &
MJD              &
Frequency        \\

mm/dd/yy         &
TDB              &
(Hz)             \\

\hline

    03/14/00     &  51617.000     &  0.1335581(10)    \\
    07/04/00     &  51729.844     &  0.1334919(16)   \\
    07/10/00     &  51735.519     &  0.1334915(22)   \\
    07/16/00     &  51741.478     &  0.133477(6)     \\
    07/23/00     &  51748.413     &  0.1334830(12)   \\
    07/30/00     &  51755.496     &  0.1334765(13)   \\
    08/04/00     &  51760.321     &  0.1334764(23)   \\
    08/12/00     &  51768.801     &  0.133454(21)    \\
    08/23/00     &  51779.522     &  0.1334655(33)   \\
    09/03/00     &  51790.574     &  0.1334566(10)   \\
    
\hline

\end{tabular}

\end{center}
\end{table}

\begin{table}[!hp]
%\vspace{1.3in}
\begin{center}
\tighten
%\footnotesize
\caption{Year 2000 pulse ephemerides for SGR~1806$-$20 from {\it RXTE} 
PCA observations.
\label{tbl:4}}
%\tablewidth{0pt}
\vspace{11pt}
%\small

\begin{tabular}{lcc}
\hline
\hline

Parameter           &
Ephemeris 2000a     &
Ephemeris 2000b     \\

\hline

Date Range  &  2000 May 01 $-$ June 24 &  2000 September 08 $-$ November 23 \\

MJD Range      &  51666.0 $-$ 51719.9  &  51795.8 $-$ 51871.6 \\ 

Epoch (MJD)                            &  51690.0  &  51840.0  \\ 

$\chi^2$/dof                           &  6.7/5    &  11.0/7   \\ 

$\nu$ (Hz)                             &  0.133516758(7)  &  0.13342617(2)  \\

$\dot{\nu}$ (10$^{-12}$ Hz s$^{-1}$)   &  -6.58(1)      &  -7.516(22)    \\

$\ddot{\nu}$ (10$^{-20}$ Hz s$^{-2}$)  &  ...      &  -1.78(21)    \\

$\nu^{(3)}$ (10$^{-26}$ Hz s$^{-3}$)   &  ...      &  8.0(28)    \\

\hline

\end{tabular}

\end{center}
\end{table}

\end{document}